\documentclass[journal,twoside,web]{ieeecolor}
\usepackage{generic}
\usepackage{cite}
\usepackage{soul}
\soulregister\cite7
\soulregister\ref7
\usepackage{amsmath,amssymb,amsfonts}
\usepackage{algorithmic}
\usepackage{graphicx}
\usepackage{textcomp}
\usepackage{subfigure}
\usepackage{amsmath, bm}
\usepackage{amsfonts,amssymb}
\usepackage[justification=centering]{caption}
\usepackage{multirow}
\usepackage{float}
\usepackage{easyReview}
\usepackage{threeparttable}
\usepackage[T1]{fontenc}
\usepackage{tcolorbox}
\usepackage{epstopdf}
\usepackage{colortbl}
\usepackage[colorlinks, bookmarksopen, bookmarksnumbered, citecolor=green, linkcolor=red, urlcolor=blue]{hyperref}
\usepackage[hyphenbreaks]{breakurl}

\def\BibTeX{{\rm B\kern-.05em{\sc i\kern-.025em b}\kern-.08em
    T\kern-.1667em\lower.7ex\hbox{E}\kern-.125emX}}
\begin{document}
\title{Progressive Dual Priori Network for Generalized Breast Tumor Segmentation}
\author{Li WANG,  Lihui WANG*, Zixiang KUAI*, Lei TANG*, Yingfeng OU, Min WU, Tianliang SHI, Chen YE, Yuemin ZHU
\thanks{This work was supported by the National Natural Science Foundation of China (Grant No.62161004), Guizhou Provincial Science and Technology Projects (QianKeHe ZK [2021] Key 002), the Nature Science Foundation of Guizhou Province (QianKeHe [2020]1Y255), Guizhou Provincial Science and Technology Projects (QianKeHe ZK [2022] 046) and Guizhou Provincial Basic Research Program(QianKeHe ZK [2023] 058). (Corresponding author: Lihui WANG, Zixiang KUAI and Lei TANG).}
\thanks{Li Wang, Lihui Wang, Yingfeng Ou and Chen YE are with Engineering Research Center of Text Computing \& Cognitive Intelligence, Ministry of Education, Key Laboratory of Intelligent Medical Image Analysis and Precise Diagnosis of Guizhou Province, State Key Laboratory of Public Big Data, College of Computer Science and Technology, Guizhou University, Guiyang, 550025, China(e-mail: wangli614079@163.com; (lhwang2@gzu.edu.cn, wlh1984@gmail.com); 2452342508@qq.com; yechenfish@163.com).}
\thanks{Zixiang KUAI is with Imaging Center, Harbin Medical University Cancer Hospital, Harbin, 150081, China(e-mail: zixiangkuai@126.com).}
\thanks{Lei TANG is with Radiology department, Guizhou Provincial People's Hospital, Guiyang, 550002, China(e-mail: tanglei7213@126.com).}
\thanks{Min WU and Tianliang SHI is with Radiology department,Tongren People's Hospital, Tongren, 554300, China(e-mail: wmzyx18@126.com;shitianliang0106@126.com).}
\thanks{Yueming ZHU is with Univ Lyon, INSA Lyon, CNRS, Inserm, CREATIS UMR 5220, U1206, F-69621, Lyon, France(e-mail: yue-min.zhu@creatis.insa-lyon.fr).}}

\maketitle

\begin{abstract}
To promote the generalization ability of breast tumor segmentation models, as well as to improve the segmentation performance for breast tumors with smaller size, low-contrast and irregular shape, we propose a progressive dual priori network (PDPNet) to segment breast tumors from dynamic enhanced magnetic resonance images (DCE-MRI) acquired at different centers. The PDPNet first cropped tumor regions with a coarse-segmentation based localization module, then the breast tumor mask was progressively refined by using the weak semantic priori and cross-scale correlation prior knowledge. To validate the effectiveness of PDPNet, we compared it with several state-of-the-art methods on multi-center datasets. The results showed that, comparing against the suboptimal method, the DSC and HD95 of PDPNet were improved at least by 5.13\% and 7.58\% respectively on multi-center test sets. In addition, through ablations, we demonstrated that the proposed localization module can decrease the influence of normal tissues and therefore improve the generalization ability of the model. The weak semantic priors allow focusing on tumor regions to avoid missing small tumors and low-contrast tumors. The cross-scale correlation priors are beneficial for promoting the shape-aware ability for irregular tumors. Thus integrating them in a unified framework improved the multi-center breast tumor segmentation performance. The source code and open data can be accessed at \url{https://github.com/wangli100209/PDPNet}.
\end{abstract}


\section{Introduction}
\label{sec:introduction}

\IEEEPARstart{B}{reast} cancer has become the most prevalent cancer worldwide \cite{bray2018global,sung2021global}, its early diagnosis and precise treatment can undoubtedly increase the survival rate. To assist the diagnosis and to further facilitate the quantitative analysis, the detailed delineation of the tumor regions is usually required.  However, due to the complexity of shapes and sizes of the breast tumors, manual annotation of the regions of interest is very time-consuming and labor-intensive, in addition, the annotation accuracy highly relies on the experiences of the experts. Therefore, investigating accurate and automatic breast tumor segmentation algorithms is of great significance.

With the successful development of deep learning models in the filed of computer vision, numerous learning-based studies have been dedicated to automatically segmenting breast tumors from mammography images \cite{li2021dual,baccouche2021connected,soulami2021breast}, ultrasound images \cite{huang2020segmentation, lei2021breast, lee2020channel} and magnetic resonance images (MRIs) \cite{zhang2022automatic, haq2022bts}, especially for dynamic contrast enhanced MRIs (DCE-MRIs). For instance,  Benjelloun et al. segmented breast tumors from DCE-MRIs with UNet and obtained a relative high mean intersection-over-union region (mIOU) value, demonstrating the feasibility of deep learning model for breast tumor segmentation \cite{benjelloun2018automated}.
To further reduce the effect of thoracic cavity region and promote the breast tumor segmentation performance, Zhang et al. used a cascade FCNet to segment the whole-breast with the pre-contrast MRIs firstly and then delineated tumors from the breast region with all the DCE-MRIs, even though it achieved a better performance, its two-stage training manner was time- and computation resource-consuming \cite{zhang2018hierarchical}.
Considering the sequential property of time-intensity curve (TIC) of DCE-MRIs and the merits of long and short-time memory (LSTM) network in exploring sequential data, Chen et al. combined LSTM and convolutional neural network (CNN) to segment breast tumors from 3D DCE-MRIs\cite{chen2018spatio}. Moreover, the kinetic  maps calculated from the DCE-MRIs can reflect pathological properties of tumors, which may provide complementary information for quantifying  the tumor regions, accordingly, Qiao et al. used both semi-quantitative parameter maps and DCE-MRIs as the input of network to predict breast tumor regions,  achieving a promising performance\cite{qiao2021three}. Although using the prior physical knowledge of DCE-MRIs can increase breast tumor segmentation accuracy, the segmentation performance for non-mass enhancement tumors with small size and irregular shapes is still not satisfactory due to the influence of background parenchymal enhancement and large variations in tumor size and shape.

To address the above-mentioned issues, Wang et al. proposed a novel tumor-sensitive synthesis module to decrease the false-positive segmentation by introducing the differential loss between true and false breast tumors \cite{wang2021breast}. Considering that the multi-perspectives information are complementary and may improve the segmentation performance, Wang et al. fused the image features of adjacent slices and combined the multiscale features to improve the segmentation accuracy \cite{wang2021mixed}. Meanwhile, given that multiscale feature representations are beneficial for capturing the details of the target and the transformer module is good at extracting global semantic information, Qin et al. introduced a multi-scale parallel convolution fusion module and a transformer module into the UNet architecture to delineate the tumor regions, which enables the model to enhance the edge and fine information \cite{qin2022joint}. To further increase the segmentation accuracy of tumors with irregular shapes, Peng et al. used the adaptive deformable convolution layer to capture the details of tumor edges\cite{peng2022lma}, Hirsch et al. trained a deep network with non-overlapping image patches to recover the tumor positional information\cite{hirsch2021radiologist}, and Zhu et al proposed a 3D affinity learning-based multi-branch ensemble network to refine the tumor and boundary voxels \cite{zhou2022three}. Even though modifying network structure and fully using kinetic parameters derived from DCE-MRI offer a promising paradigm for breast tumor segmentation, most of the existing models are two-stage learning-based, which requires to segment breast region first and then the tumors, affecting not only the annotation but also the computing efficiency. In addition, most of them did not consider the multicenter segmentation problems and have poor generalizability across different image domains.

Recently, some semi-supervised \cite{you2022simcvd,luo2022semi,wang2022semi} and domain adaptation strategies \cite{xia2020uncertainty,hu2022domain,chen2023reconstruction} were proposed to deal with the influence of domain shift on the generalization ability of deep learning models. Semi-supervised strategies, such as pseudo label and consistency regularization, intend to extract useful information from unlabeled data to increase the generalization ability of the model, while the domain adaptation tries to reduce the distribution divergence between different domains by aligning their features in manifold space or extracting domain-invariant features. Although these strategies can generalize well, requiring to access the target data (test data) during the model training makes them impractical since such data is usually unseen in advance. To promote the generalization ability of segmentation models without using target data, as well as improve the segmentation performance for breast tumors with irregular shape, non-enhanced signal and small size, we propose a progressive dual priori network (PDPNet). Specifically, the main contributions of this work are as follows:

1) We presented  a coarse-segmentation guided localization module that  is used to identify the tumor-containing regions. It can provide a weak-supervised information for the subsequent segmentation task and  improve the generalization ability of the segmentation model by reducing the influence of  noise and heterogeneous background intensities on the feature representation.

2) We designed a dual priori based segmentation module, which refines gradually segmentation results by predicting multi-scale tumor masks with priori-weighted pyramid features. The dual priori module enables the model focus on the tumor regions at different scales, where weak semantic maps are beneficial for distinguishing tumors with small size and cross-scale spatial correlation maps are useful for refining the tumor shapes.

3) We built a paradigm to promote the generalization ability of breast tumor segmentation models without using target data, and demonstrated that the proposed method can segment breast tumors successfully from DCE-MRIs acquired at different centers.

\section{Methods}
\label{sec:Materials and Methods}

\begin{figure*}[!h]
	\centering
	\vspace{0cm}
	\setlength{\abovecaptionskip}{0cm}
	\setlength{\belowcaptionskip}{0cm}
	\includegraphics[width=0.8\textwidth]{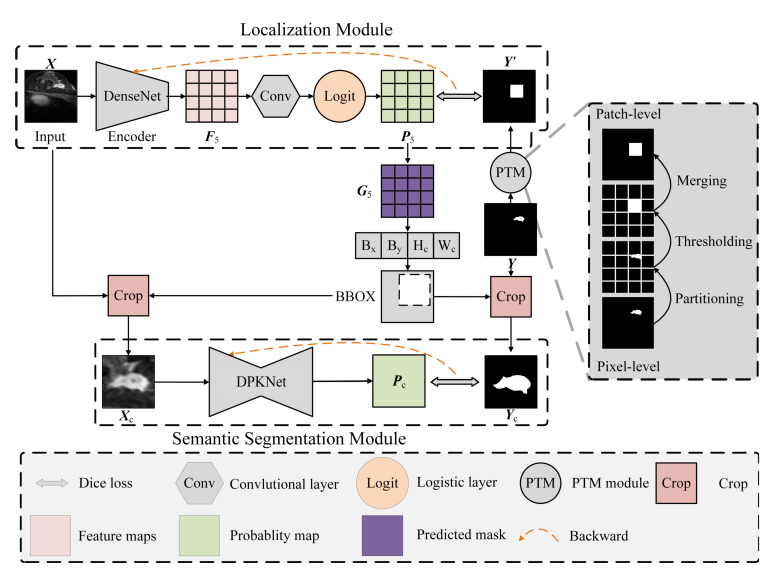}
	\captionsetup{justification=justified}
	\caption{Overall structure of PDPNet. It consists of a localization module which is used to limit the influence of the background information, and a dual prior knowledge based segmentation network (DPKNet) which is responsible for segmenting the breast tumors in a progressive manner.}
	\label{fig:1}
\end{figure*}

The detailed structure of PDPNet is given in Fig. \ref{fig:1}. The input image $\bm {X}$ is fed into the localization module, which outputs the size $(H_c,W_c)$ and top-left corner coordinate $(B_x,B_y)$ of a bounding box. The bounding box is then used to derive the cropped input $\bm{X_c}$ of dual priori knowledge based network (DPKNet). After that, the original label map $\bm{Y}$ is partitioned into a smaller label map $\bm{Y^\prime}$ through a partition-thresholding-merging (PTM) operation, and is also cropped with bounding box to generate $\bm{Y_c}$. Taking $\bm{Y^\prime}$ and $\bm{Y_c}$ as the learning target respectively, the localization module and DPKNet can be trained. The detailed process will be elaborated in the following subsections. 

\subsection{Localization Module}

In the localization module, the original image ${\bm{X}}$ with a size of 128 $\times$ 128 passes through a DenseNet121 to extract the feature representation maps $\bm{F}$ of size 4 $\times$ 4. After a $1\times 1$ convolutional layer and a logistic layer, a probability map $\bm{P}$ is generated to predict whether it contains tumor in each patch (each pixel  in a $4 \times$ 4 probability map corresponds to an image patch of size $32 \times 32$). Since the size of the original annotation map is 128 $\times$ 128, to get the ground-truth for this patch-level prediction, a PTM operation on original label map $\bm{Y}$ is used, as illustrated in Fig. \ref{fig:1}. In the partition step, the ratio of tumor region to each image patch is calculated by:
\begin{equation}
	\label{eq:1}
	r_{i,j}= \frac{{\sum\limits_{u \in \Omega (i,j)} {{y_u}} }}{{Area(\Omega (i,j))}} \quad \text {with }  i,j=1,2...4,
\end{equation}
where  $\Omega (i,j)$ represents the annotation image patch corresponding to the pixel $(i,j)$ in probability map $\bm{P}$,  $y_u\in\{0,1\}$ indicates the label for a given pixel in the patch  $\Omega (i,j)$, and  the $Area(\cdot)$ denotes the area of patch region. To determine whether the patch  $\Omega (i,j)$ contains the tumor, a thresholding operation is performed on $r_{i,j}$ to generate the patch-level label $\bm{Y}^{\prime}_{i,j}$,
\begin{equation}
	\label{eq:2}
	\mathrm{y}=\left\{
	\begin{array}{ll}
		1 & \text { if } \quad  r_{i, j}>\frac{\sum_{p=1}^W \sum_{q=1}^H y_{p, q}}{H \times W} \\
		0 & \text { Otherwise }
	\end{array},
	\right.
\end{equation}
in which,  $y_{p,q}\in\{0,1\}$ is the label of  pixel $(p,q)$ in raw annotation image $\bm{Y}$,  $H$ and $W$ denote the height and width of $\bm{Y}$. It means that if the tumor ratio in a patch is larger than the ratio of whole tumor region to the original image size, the patch-level label is noted as 1. Merging all the patch-level labels results in the patch-level annotation map $\bm{Y^\prime}$. Using such patch-level annotation map  $\bm{Y^\prime}$ as the target to train the localization module, the tumor region mask $\bm{G}$ can be obtained by thresholding the output probability map $\bm{P}$ with a cut of 0.5. In $\bm{G}$, the voxels predicted as tumor region are labeled with 1 and the other voxels are labeled with 0. Based on the predicted tumor mask $\bm{G}$, we can calculate the height $H_c$ and width $W_c$ of tumor bounding box with $m\times32$ and $n\times 32$, as well as the top-left corner coordinate of the box with $B_x\times32$ and $B_y\times 32$, where $B_x$ and $B_y$ correspond to the voxel index of the first occurrence of label 1  when traversing $\bm{G}$ from left to right and from top to bottom respectively, $m$ and $n$ are the maximum number of voxels labeled as 1 along row and column directions of $\bm{G}$. Accordingly, the centroid  $(C_x, C_y)$ of the tumor bounding box is calculated as $(B_x*32+W_c/2, B_y*32+H_c/2)$. Considering that the prediction result of the localization network may bring the bias in centroid and size of the tumor regions, if the bounding box is strictly defined, such bias may result in a predicted bounding box containing incomplete tumor. To deal with this issue, we expand the height and width of the tumor bounding box. Meanwhile, to ensure that the cropped image fed into the segmentation module have a regular shape, we let the expanded height and width of the bounding box be the same, noted as $S_c$,
\begin{equation}
	\label{eq:3}
	S_c =\max ({H_c},{W_c})+ \left\lfloor { \frac{{\min (H,W) - \max ({H_c},{W_c})}}{2}} \right\rfloor 
\end{equation}
where $\lfloor \cdot \rfloor$ denotes round down function. The vector $(C_x, C_y, S_c, S_c)$ is finally used to crop the image.

\subsection{Dual Priori Knowledge based Segmentation Network }

\begin{figure*}[!h]
	\centering
	\vspace{0cm}
	\setlength{\abovecaptionskip}{0cm}
	\setlength{\belowcaptionskip}{0cm}
	\includegraphics[width=0.8\textwidth]{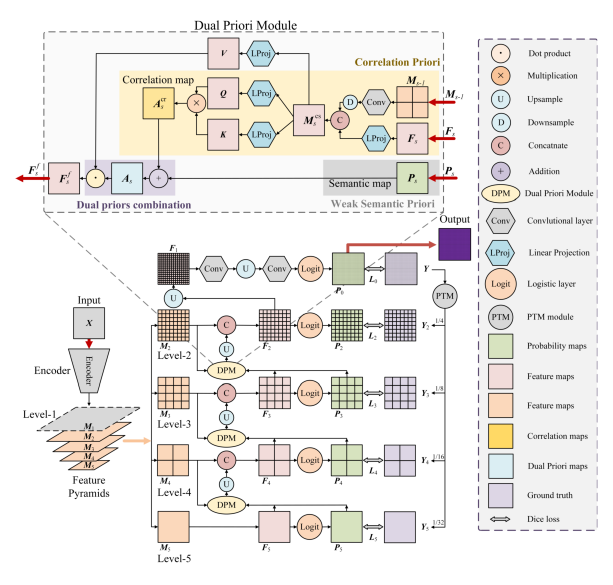}
	\captionsetup{justification=justified}
	\caption{Dual priori knowledge based network (DPKNet). It has an encoder-decoder architecture, with a dual prior module (DPM) being used between the low-level encoded features $M_{s-1}$ and high-level decoded features $F_s, (s={3,4,5})$ to promote the segmentation performance.}
	\label{fig:2}
\end{figure*}

The objective of semantic segmentation network is to learn a mapping function which can convert the input image $\bm{X}$ into the corresponding  label  map   $\bm{Y}$ in a patch-to-voxel manner, which means that the value of each voxel  $Y_{i,j}$ in label map is determined by the features of one local image patch $\Omega_x (i,j)$ around the voxel $(i,j)$, written as, 
\begin{equation}
	\label{eq:4}
	\begin{array}{c}
		Y_{i,j}=f(\theta; g(\phi;\bm X)_{\Omega_x(i,j)}))	\\
	\end{array}
\end{equation}
where $ g(\phi;\bm X)$ represents a feature extraction module with the learnable parameters $\phi$, and  $ g(\phi;\bm X)_{\Omega_x(i,j)}$  denotes the features corresponding to the patch ${\Omega_x(i,j)}$. Considering that the patch size used for predicting the label of  $Y_{i,j}$ is determined by the semantic feature level,  higher semantic level corresponds to the bigger patch size and less dense prediction. It is relative easy to determine which pathes contain tumor and accordingly promotes the generalization ability of the prediction model. However, the localization module is prone to loosing the location and shape information of the tumor and therefore is not able to segment the tumor details. To make a compromise between segmentation performance on multicenter dataset and details segmentation accuracy, we propose a dual priori knowledge based network (DPKNet), as illustrated in Fig. \ref{fig:2}, which uses  weak semantic priors at different levels and cross-scale correlation priors to guide the segmentation.

DPKNet takes the cropped image $\bm{X_c}$ as input, and uses the DenseNet as encoder to extract multi-level feature maps $\bm{M_s}$ ($s=1,2,3,...,5$).

From these pyramid feature maps, weak semantic prior at each scale is estimated by performing the patch-level tumor segmentation task. In other words, we first apply the PTM operation on the cropped label map $\bm{Y_c}$ to get patch-level labels for different scales, noted as $\bm{Y}_s$, and then using the pyramid features to predict the tumor probability maps $\bm{P}_s$ (also called weak semantic priors) at different scales through convolution, BN and Sigmoid layers (Fig. \ref{fig:2}).

Since low-level feature maps (with bigger size)  $\bm{M}_{s-1}$ derived from the encoder contain more texture information, while high level (with smaller size) feature maps $\bm{F}_{s}$ extracted from the decoder comprise more semantic information, using the cross-scale correlation between them can provide generalized shape-aware details of the tumor. Meanwhile, the high-level weak semantic maps  $\bm{P}_s$ can predict more accurately the tumor location. Combining such semantic priori with the correlation prior may enhance the feature representation ability about the tumor, thus improving the generalization performance of the model. Based on such assumption, we designed the dual prior module (DPM), as shown in Fig. \ref{fig:2}. It takes the low-level feature maps $\bm{M}_{s-1}$, high-level feature maps $\bm{F}_s$, and the weak semantic map $\bm{P}_s$ as the input. To derive the cross-scale correlation prior, the self-attention between the low- and high-level feature maps is calculated. Specifically, the low-level feature maps $\bm{M}_{s-1}$ pass through the convolution and downsampling operations and the high-level feature maps $\bm{F}_{s}$ are linearly projected to make them have the same size. After concatenating them the cross-scale feature maps $\bm{M}_s^{cs}$ were derived, 
\begin{equation}
	\label{eq:7}
	\bm{M}_s^{cs} = Cat\left[ \downarrow Conv(\bm{M}_{s - 1}),LProj(\bm{F}_s)\right] 
\end{equation}
where $\downarrow $ and $Conv$ represent down-sampling and convolution operations respectively, $LProj$ means the linear projection, and $Cat$ denotes the concatenation. Accordingly, the cross-scale correlation prior $\bm{A}_s^{cr}$ is calculated through the self-attention mechanism, formulated as:
\begin{equation}
	\label{eq:8}
	\begin{aligned}
		\bm{A}_s^{cr} & = \sigma (Q \times K) \\
		& =\sigma (LProj_q(\bm{M}_s^{cs}) \times LProj_k(\bm{M}_s^{cs}))\\ 
	\end{aligned}
\end{equation}
where ${\sigma ( \cdot )}$ denotes the sigmoid function, ''$\times$'' means outer product, $LProj_q$ and $LProj_k$ are the linear projections to derive query $Q$ and key $K$ vectors, respectively. 

To further enhance the feature representation with location and shape-aware information, the weak semantic prior $\bm{P}_s$ and correlation prior $\bm{A}_s^{cr}$ are used simultaneously as the attention score to re-weight the feature map, result in the output of DPM $F_s^f$, 
\begin{equation}
	\label{eq:9}
	\begin{aligned}
		\bm{F}_s^{f} & = (\bm{A}_s^{cr} + {\bm{P}_s}) \cdot V\\
		& = (\bm{A}_s^{cr} + {\bm{P}_s})\cdot (LProj_v(\bm{M}_s^{cs}))\\
	\end{aligned}
\end{equation}
where ''$\cdot$'' denotes the inner product, $LProj_v$ indicates the linear projection for generating the value vector $V$.

In addition, to ensure the segmentation accuracy, skip connection between low-level features $\bm{M}_s$ and the weighted high-level semantic features $\bm{F}_s^f$ is also used, followed by a convolutional block; the decoded feature maps $\bm{F}_{s-1}$ are formulated as:
\begin{equation}
	\label{eq:10}
	\begin{aligned}
		\bm{F}_{s-1} &= Conv(Cat[{\bm{M}_{s - 1}}, \uparrow \bm{F}_s^f]) \quad s=2,3,4,5 \\
		\bm{F}_{5} &= Conv(\bm{M}_5) \\
	\end{aligned}
\end{equation}
where ''$\uparrow $'' means the upsampling operator with upscale of 2. Using the decoded feature maps at different levels $\bm{F}_s$, the weak segmentation priors $\bm{P}_s$ can be predicted with a logistic layer. Note that, since the low-level feature maps $\bm{M}_1$ of the encoder contain fewer semantic information, the corresponding cross-correlation prior $A_2^{cr}$ calculated from $\bm{M}_1$ and $\bm{F}_2$ may present a large bias, which may influence the final segmentation result. Accordingly, in this work, $M_1$ and $F_1$ were not involved in the final prediction (as illustrated in Fig.\ref{fig:2}). Instead, by minimizing the losses between $\bm{P}_s$ and $\bm{Y}_s (s=2,3,4,5)$ at each level, the semantic feature map $\bm{F}_2$ can be gradually refined, based on which the final segmentation map $\hat{\bm{Y}}_c$ is inferred through upsampling, convolution and logistic layers, detailed as:
\begin{equation}
	\label{eq:9}
	\hat{\bm{Y}}_c= logit(Conv(\uparrow Conv(\uparrow \bm{F}_{2})))
\end{equation}

\subsection{Loss Functions}
All the tasks, including the tumor localization, the patch-level and pixel-level segmentation are trained simultaneously in an end-to-end manner, but it is worth noting that the localization network and DPKNet do not propagate the gradient to each other. Since the localization network used in this work is in fact realized by a corse patch segmentation, we use Log-Cosh Dice loss [20] as objective function to optimize its parameters: 
\begin{equation}
	\label{eq:9}
	{loss_{loc}} = \log (cosh({l_{Dice}}(\bm{P},\bm{Y^{\prime}})))
\end{equation}
where $\bm{P}$ and $\bm{Y^{\prime}}$ represent the output probability map and the ground-truth label of localization network, respectively, ${l_{Dice(\bm{\cdot}, \bm{\cdot})}}$ indicates the Dice loss. Given any two images $I^1$ and $I^2$, the dice loss between them is defined by: 
\begin{equation}
	\label{eq:10}
	{l_{Dice(I^1, I^2)}} = 1 - \frac{2{\sum\limits_j^W {\sum\limits_i^H {{I^1_{i,j}}{I^2_{i.j}}} } }}{{\sum\limits_j^W {\sum\limits_i^H {{I^1_{i,j}}} }  + \sum\limits_j^W {\sum\limits_i^H {{I^2_{i,j}}} } }}
\end{equation}
As to the DPKNet, its loss is composed by the multi-scale patch-level and pixel-level segmentation losses, written as: 
\begin{equation}
	\label{eq:11}
	\begin{aligned}
		loss_{seg} &= \sum \limits_{s=2}^5 {\log (cosh({l_{Dice}}(\bm{P}_s,\bm{Y}_s)))} \\
		& + \log(cosh(l_{Dice}(\hat{\bm{Y}}_c,\bm{Y}_c)))	
	\end{aligned}
\end{equation}

\section{Experiments}
\label{sec:Experiments}
\subsection{Data Description and preprocessing}

The DCE-MRI dataset of breast tumor utilized in this study originated from six distinct centers, denoted as HUM cohort, GPH cohort, HUM-S cohort, TPH cohort, ISPY-G cohort and ISPY-S cohort (\url{https://www.cancerimagingarchive.net/collection/ispy1}) \cite{newitt2016multi, hylton2016neoadjuvant, clark2013cancer}, respectively. The HUM cohort included 3D DCE-MRI images of 243 breast cancer patients, 70\% of which was used as the training set, while the remaining 30\% served as the validation set. The other five cohorts were all considered as the testing sets. Besides the publicly available dataset, all the experimental protocols have been approved by the ethics committee of Guizhou Provincial People's Hospital (No. (2023)015). The tumor regions in all these datasets were delineated by two experienced radiologists. Since the datasets used in this work come from different centers, they were acquired with different  MRI scanners, different sequences, and different imaging parameters. In addition, for each patient, the number of contrast-enhanced MRI examinations and the time resolution of DCE-MRIs, as well as the breast tumor receptor subtypes are different. The detailed DCE-MRI acquisition parameters and demographic information for each cohort is presented in Table \ref{table:1}.

\begin{table*}[htb]
	\centering
	\captionsetup{labelfont={color=blue} }
	\caption{Detailed DCE-MRI acquisition parameters and demographic information of different cohorts.}
	\resizebox{.8\textwidth}{!}{					
		\begin{tabular}{lccccccc}		
			\hline			
			& \textbf{Training set} & \textbf{Validation set}                               & \multicolumn{5}{c}{\textbf{Testing sets}} \\			
			& HUM (70\%)     & HUM (30\%)   & GPH    &HUM-S    & ISPY-G    &ISPY-S    &TPH  \\  \hline			
			\textbf{Paitents} & 170     & 73    & 72    & 15    & 65    & 24    & 25 \\			
			\textbf{Total Slices}   & 11185    & 2719    & 2086    & 270     & 981    & 238      &200 \\			
			\textbf{MRI scanner}    & \multicolumn{2}{c}{3T Philips (Ingenia)}    & 3T Siemens    & 3T GE    & 3T GE    &  3T Siemens    & 3T Philips \\		
			\textbf{Sequence}    & \multicolumn{2}{c}{e-THRIVE}    & GRE    & T1-SPGRE    & Scientific    & Scientific    & GRE \\			
			\textbf{Pre-contrast scans}    & \multicolumn{2}{c}{1}    & 1    & 1    & 1    &1    &1 \\			
			\textbf{Post-contrast scans}    & \multicolumn{2}{c}{5}    & 1$\sim$6    & 4    & 2    &2    & 1 \\
			\textbf{TR ($ms$)}    & \multicolumn{2}{c}{4.8}    & 4.67    & 4.4    & 8.4   &3    & 3.6 \\	
			\textbf{TE ($ms$)}    & \multicolumn{2}{c}{2.1}    & 1.6   & 2.1    & 4.2    &4    & 1.8 \\
			\textbf{Flip angle ($^\circ$)}    & \multicolumn{2}{c}{12}    & 10   & 12     & 20    &45    & 11 \\						
			\textbf{In-plane resolution($mm^2$)}    & \multicolumn{2}{c}{$0.45 \times 0.45$}    & $1 \times 1$    & $0.625 \times 0.625$    & $0.75 \times 0.75$    & $ 0.75 \times 0.75 $    & $ 0.625 \times 0.625 $ \\			
			\textbf{Slice thickness($mm$)}    & \multicolumn{2}{c}{0.75}    & 1.5    & 0.5    & 2    & 2    & 3 \\			
			\textbf{Matrix size}    & \multicolumn{2}{c}{$\geq 512 \times 512$}    & $\geq 300 \times 300$    & $\geq 512 \times 512$    & $256 \times 256$    & $256 \times 256$    & $\geq 512 \times 512$ \\
			\textbf{Time resolution($s$)}    & \multicolumn{2}{c}{0.47$\sim$0.94}    & 0.50$\sim$1.50    & -    & 5.50$\sim$6.00    & 3.75$\sim$7.79    & - \\
			\textbf{Slice number}    & \multicolumn{2}{c}{$\geq 128$}    & $\geq 128$    & $\geq 292$    & 60    & 64    & $\geq 36$ \\			
			\textbf{Age}      & & & & & & & \\			
			\textless{}40     & 40    & 10    & 16    & -    & 15    & 3    & 6 \\			
			\textless{}50     & 56    & 36    & 28    & -    & 26    & 7    & 6 \\			
			\textgreater{}=50 & 74    & 27    & 28    & -    & 24    & 14   & 13 \\
			\textbf{ER}      & & & & & & & \\
			Positive    &85 &23 &57 &- &38 &13 &- \\
			Negative    &22 &9 &15 &- &27 &11 &- \\
			Unknown    &63 &41 &0 &- &0 &0 &- \\
			\textbf{PR}      & & & & & & & \\
			Positive    &58 &17 &59 &- &32 &12 &- \\
			Negative    &42 &21 &13 &- &33 &12 &- \\
			Unknown    &70 &35 &0 &- &0 &0 &- \\
			\textbf{HER-2}      & & & & & & & \\
			Positive    &96 &37 &64 &- &19 &7 &- \\
			Negative    &1 &1 &8 &- &46 &17 &- \\
			Unknown    &73 &35 &0 &- &0 &0 &- \\
			\hline			
		\end{tabular}			
	}	
	\begin{tablenotes}
		\footnotesize
		\item [*] GRE: gradient recalled echo sequence; Scientific: scientific mode sequence; ER: estrogen receptor; PR: Progesterone receptor; HER-2: Human epidermal growth factor receptor-2.
	\end{tablenotes}
	\label{table:1}
\end{table*}

To decentralize the data at the patient level, zero-mean normalization was firstly implemented, and then all the images were resized as 128$\times$128. To avoid the overfitting, the data augmentations strategies, such as image scaling, translations, and rotations are used during the training. 

\subsection{Experimental setup}

To validate the effectiveness of the proposed method, we compared it with several state-of-the-art (SOTA) models, including Unet3+\cite{huang2020unet}, UTNet\cite{gao2021utnet}, DUNet\cite{jha2020doubleu}, LEDNet \cite{wang2019lednet}, BiSeNetV2\cite{yu2021bisenet}, AttUnet\cite{oktay2018attention}, DeeplabV3\cite{chen2017rethinking},AAUnet\cite{chen2022aau}, EUnet\cite{chowdary2023eu}, UNest\cite{yu2023unest}, and nnUnet\cite{isensee2021nnu}. All the models were trained on NVIDIA Tesla P40 GPUs and tested on NVIDIA GeForce GTX 1660Ti GPUs using the Pytorch framework, batch size = 32 and epoch=500. To fairly compare the different model structures or strategies and avoid the influence of optimizers, Adam optimizer was used in all the comparison models since it can make them achieve their best performance, in which the learning rate was set as $\eta  = 0.0001$, and momentum parameters were ${\beta _1} = 0.9$ and ${\beta _2} = 0.999$. However, considering that the multiple losses used in the proposed model constraint each other which makes the gradient become very small, in this case, using ADAM optimizer with learning rate decreased gradually may influence the convergence speed. Accordingly, in our PDPNet, the SGD with a large learning rate ($\eta  = 0.001$) rather than ADAM was used to optimize the model parameters.

\subsection{Evaluation metrics}
Dice similarity coefficient (DSC), Hausdorff distance 95\%(HD95), sensitivity(SEN), and specificity (SPE)  were used to quantitatively evaluate the segmentation performance of each model. Among them, DSC measures the regional overlap score between predictions and labels, defined as:
\begin{equation}
	\label{eq:11}
	\text{DSC} = \frac{2{\sum\limits_{j=1}^W {\sum\limits_{i=1}^H {{g_{i,j}}{y_{i,j}}} } }}{{\sum\limits_{j=1}^W {\sum\limits_{i=1}^H {{g_{i,j}}} }  + \sum\limits_{j=1}^W {\sum\limits_{i=1}^H {{y_{i,j}}} } }}
\end{equation}
where $g_{i,j}$ and $y_{i,j}$ denote respectively the values of predicted tumor mask and ground-truth label at the voxel $(i,j)$. DSC ranges from 0 to 1;  higher DSC indicates better segmentation performance. 

HD95 reflects similarity in shape and edge between the predicted mask and label, formulated as:

\begin{equation}
	\label{eq:13}
	\text{HD95} = \max \{ \mathop {CI_{95} }\limits_{y \in Y} \{ \mathop {min}\limits_{g \in G} \{ \left\| {y - g} \right\|\} \} ,\mathop {CI_{95} }\limits_{g \in G} \{ \mathop {min}\limits_{y \in Y} \{ \left\| {g - y} \right\|\} \} \}
\end{equation}

where $CI_{95}\{\cdot\}$ denotes the 95\% confidence interval, $\left| \cdot \right|$ means the number of non-zero elements in the set, $\Vert \cdot \Vert$ represents L2 norm distance. HD95 has no upper limit, the lower the HD95, the higher the shape and edge similarity become. 

SEN and SPE evaluate the ability of a model to correctly identify tumor and non-tumor pixels, respectively, written as:
\begin{equation}
	\tag{15}
	\begin{aligned}
		SEN &= \frac{{\sum\limits_{j=1}^W {\sum\limits_{i=1}^H {{g_{i,j}}{y_{i,j}}} } }}{{\sum\limits_{j=1}^W {\sum\limits_{i=1}^H {{y_{i,j}}} } }} \\
		SPE &= \frac{{\sum\limits_{j=1}^W {\sum\limits_{i=1}^H {{\left( 1-g_{i,j} \right)}{\left( 1-y_{i,j} \right)}} } }}{{\sum\limits_{j=1}^W {\sum\limits_{i=1}^H {{\left( 1-y_{i,j} \right)}} }}}
	\end{aligned}
\end{equation}

The values of SEN and SPE range from 0 to 1. Higher SEN/SPE means more tumor/non-tumor pixels are correctly predicted by model.

\section{Results}
\label{sec:Results}

\subsection{Comparison with SOTA methods}

\begin{figure*}[ht]
	\setlength{\abovecaptionskip}{0cm}
	\setlength{\belowcaptionskip}{0cm}
	\subfigure[Visual comparisons on single-center cohort]{
		\setlength{\abovecaptionskip}{0cm}
		\qquad \includegraphics[width=0.9\textwidth]{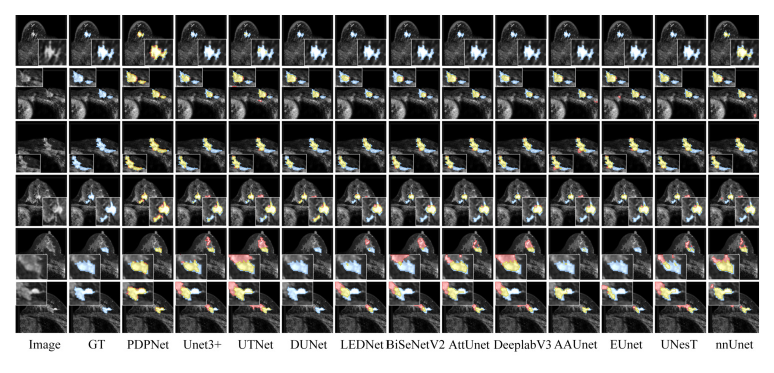}
		\label{fig:3a}
	}
	\subfigure[Visual comparisons on multicenter cohort]{
		\setlength{\abovecaptionskip}{0cm}
		\qquad \includegraphics[width=0.97\textwidth]{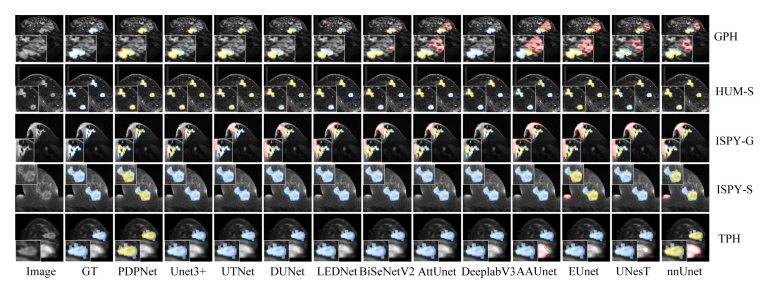}
		\label{fig:3b}
	}
	\captionsetup{justification=justified}
	\caption{Visual comparisons among various methods on validation set (a) and multi-center test sets (b). The original images and the ground truth tumor masks are presented in the first and second columns, respectively. In each sub-figure, the ground-truth tumor regions are marked in blue, the predicted tumor regions are highlighted in red and the overlap regions between segmented results and ground truth are shown in yellow. For better visualization, the local segmented tumor details are zoomed in and shown in the bottom.}
	\label{fig:3}
\end{figure*}

The segmentation results of several representative samples on validation (a) and multi-center test (b) sets obtained with different models are shown in Fig. \ref{fig:3}. It can be seen that PDPNet outperforms the other methods on validation set (Fig. \ref{fig:3a}), with the tumor regions as well as the boundaries and shapes conformed to the ground-truth well, while in the other methods, there are too much false tumors (backgrounds were taken as tumors, in red regions) or false backgrounds (tumors were predicted as backgrounds, blue regions). On the multi-center test sets, PDPNet is still able to predict tumors more precisely than the other methods, especially for the tumors with smaller size (HUM-S cohort), irregular shapes (GPH cohort) and similar intensity to the normal tissues (ISPY-S, ISPY-G, and TPH cohorts). Although EUNet and nnUNet can distinguish some of these tumors, they are easy to generate false tumors, demonstrated by the red regions in ISPY-S ISPY-G, and TPH cohorts.

\begin{table*}[htbp]
	\centering
	\renewcommand{\arraystretch}{1.0}
	\captionsetup{labelfont={color=blue}}
	\caption{Quantitative evaluation metrics (95\% confidence interval) of different segmentation methods on both validation and multi-center test sets.}
	\label{table:s3}
	\begin{threeparttable}
		\vspace{0cm}
		\setlength{\abovecaptionskip}{0cm}
		\setlength{\belowcaptionskip}{0cm}		
		\resizebox{.85\paperwidth}{!}{
			\begin{tabular}{cccccccccc}
				\hline
				\multirow{2}{*}{Metrics}    &\multirow{2}{*}{Methods}    & Validation set    &\multicolumn{6}{c}{Multi-center test sets} &\multirow{2}{*}{P-value} \\ \cline{4-9}
				&    & HUM    &GPH    &HUM-S    &ISPY-G    &ISPY-S    &TPH    &Mean & \\ \hline
				\multirow{13}{*}{DSC}     &PDPNet    &\textbf{90.4 ${\pm}$ 0.6}    &\textbf{71.3 ${\pm}$ 1.3}    &76.0 ${\pm}$ 3.4    &\underline{61.2 ${\pm}$ 2.0}    &\underline{43.2 ${\pm}$ 5.2}    &\textbf{65.8 ${\pm}$ 3.5}    &\textbf{63.5 ${\pm}$ 15.7}    & -  \\
				&UNet3+(2020)    &\underline{90.0 ${\pm}$ 0.9}    &68.3 ${\pm}$ 1.4    &67.7 ${\pm}$ 4.4    &56.4 ${\pm}$ 2.1    &31.6 ${\pm}$ 5.1    &\underline{65.3 ${\pm}$ 4.8}    &57.9 ${\pm}$ 19.2    & \textless{0.01}  \\
				&UTNet(2021)    &85.8 ${\pm}$ 1.1    &66.9 ${\pm}$ 1.5    &\underline{78.1 ${\pm}$ 3.4}    &51.6 ${\pm}$ 2.3    &29.0 ${\pm}$ 5.1    &58.2 ${\pm}$ 5.4    &56.7 ${\pm}$ 22.9    & \textless{0.01}  \\
				&DUNet(2020)    &89.3 ${\pm}$ 1.0    &67.1 ${\pm}$ 1.6    &75.9 ${\pm}$ 3.7    &57.8 ${\pm}$ 2.1    &19.8 ${\pm}$ 4.3    &61.9 ${\pm}$ 5.1    &56.5 ${\pm}$ 26.8    & \textless{0.01}  \\
				&LEDNet(2019)    &86.4 ${\pm}$ 0.9    &68.4 ${\pm}$ 1.3    &70.1 ${\pm}$ 4.3    &45.6 ${\pm}$ 2.2    &12.1 ${\pm}$ 3.7    &48.4 ${\pm}$ 6.2    &48.9 ${\pm}$ 29.1    & \textless{0.01}  \\
				&BiSeNetV2(2021)    &86.5 ${\pm}$ 0.9    &68.8 ${\pm}$ 1.3    &67.4 ${\pm}$ 4.7    &60.2 ${\pm}$ 1.8    &32.7 ${\pm}$ 4.9    &63.6 ${\pm}$ 4.5    &58.5 ${\pm}$ 18.4    & \textless{0.01}  \\
				&AttUnet(2018)    &84.3 ${\pm}$ 1.2    &66.8 ${\pm}$ 1.5    &70.1 ${\pm}$ 4.5    &57.4 ${\pm}$ 2.2    &26.8 ${\pm}$ 5.1    &57.9 ${\pm}$ 6.0    &55.8 ${\pm}$ 21.3    & \textless{0.01}  \\
				&Deeplabv3(2017)    &84.2 ${\pm}$ 1.0    &65.9 ${\pm}$ 1.4    &71.8 ${\pm}$ 3.9    &52.8 ${\pm}$ 2.0    &28.2 ${\pm}$ 4.6    &56.1 ${\pm}$ 5.5    &55.0 ${\pm}$ 20.8    & \textless{0.01}  \\
				&AAUnet(2022)    &80.5 ${\pm}$ 1.3    &56.4 ${\pm}$ 1.7    &64.8 ${\pm}$ 4.6    &44.1 ${\pm}$ 2.1    &27.4 ${\pm}$ 3.6    &60.5 ${\pm}$ 4.4    &50.6 ${\pm}$ 18.8    & \textless{0.01}  \\
				&EUnet(2023)    &87.7 ${\pm}$ 1.0    &\underline{70.3 ${\pm}$ 1.2}    &\textbf{78.2 ${\pm}$ 3.3}    &59.7 ${\pm}$ 1.9    &37.3 ${\pm}$ 4.7    &56.3 ${\pm}$ 5.6    &\underline{60.4 ${\pm}$ 19.3}    & \textless{0.01}  \\
				&UNesT(2023)    &80.2 ${\pm}$ 1.4    &63.7 ${\pm}$ 1.4    &66.4 ${\pm}$ 4.3    &55.5 ${\pm}$ 1.9    &26.5 ${\pm}$ 4.3    &62.5 ${\pm}$ 4.2    &54.9 ${\pm}$ 20.3    & \textless{0.01}  \\
				&nnUnet(2021)    &82.8 ${\pm}$ 0.9    &69.6 ${\pm}$ 1.0    &74.3 ${\pm}$ 2.8    &\textbf{64.3 ${\pm}$ 1.5}    &\textbf{43.3 ${\pm}$ 4.8}    &49.6 ${\pm}$ 3.7    &60.2 ${\pm}$ 16.4    & \textless{0.01}  \\
				\hline
				\multirow{13}{*}{SEN}    &PDPNet    &\textbf{92.7 ${\pm}$ 0.7}    &\textbf{80.6 ${\pm}$ 1.4}    &\underline{84.4 ${\pm}$ 3.6}    &70.0 ${\pm}$ 2.2    &44.3 ${\pm}$ 5.6    &\underline{89.8 ${\pm}$ 4.1}    &\underline{73.8 ${\pm}$ 22.4}    & -  \\
				&UNet3+(2020)    &\underline{88.4 ${\pm}$ 1.0}    &68.5 ${\pm}$ 1.5    &71.9 ${\pm}$ 4.5    &66.0 ${\pm}$ 2.3    &31.4 ${\pm}$ 5.1    &82.4 ${\pm}$ 5.4    &64.1 ${\pm}$ 23.9    & \textless{0.01}  \\
				&UTNet(2021)    &86.9 ${\pm}$ 1.1    &70.7 ${\pm}$ 1.6    &82.6 ${\pm}$ 3.4    &59.6 ${\pm}$ 2.5    &30.7 ${\pm}$ 5.5    &70.0 ${\pm}$ 6.8    &62.7 ${\pm}$ 24.4    & \textless{0.01}  \\
				&DUNet(2020)    &87.5 ${\pm}$ 1.0    &68.6 ${\pm}$ 1.7    &77.0 ${\pm}$ 3.9    &67.3 ${\pm}$ 2.2    &19.3 ${\pm}$ 4.4    &79.8 ${\pm}$ 5.9    &62.4 ${\pm}$ 30.7    & \textless{0.01}  \\
				&LEDNet(2019)    &83.4 ${\pm}$ 0.9    &70.6 ${\pm}$ 1.4    &69.6 ${\pm}$ 4.4    &52.9 ${\pm}$ 2.6    &11.9 ${\pm}$ 3.7    &56.0 ${\pm}$ 7.0    &52.2 ${\pm}$ 29.7    & \textless{0.01}  \\
				&BiSeNetV2(2021)    &85.1 ${\pm}$ 0.9    &74.5 ${\pm}$ 1.4    &67.4 ${\pm}$ 4.8    &75.2 ${\pm}$ 1.7    &32.5 ${\pm}$ 5.0    &82.1 ${\pm}$ 5.3    &66.3 ${\pm}$ 24.4    & \textless{0.01}  \\
				&AttUnet(2018)    &82.7 ${\pm}$ 1.3    &69.4 ${\pm}$ 1.5    &71.5 ${\pm}$ 4.6    &67.4 ${\pm}$ 2.3    &25.4 ${\pm}$ 4.9    &70.5 ${\pm}$ 7.1    &60.8 ${\pm}$ 24.7    & \textless{0.01}  \\
				&Deeplabv3(2017)    &81.5 ${\pm}$ 1.0    &65.8 ${\pm}$ 1.5    &70.3 ${\pm}$ 4.1    &58.1 ${\pm}$ 2.2    &28.4 ${\pm}$ 4.9    &59.6 ${\pm}$ 6.3    &56.4 ${\pm}$ 20.4    & \textless{0.01}  \\
				&AAUnet(2022)    &80.2 ${\pm}$ 1.4    &62.4 ${\pm}$ 1.8    &67.1 ${\pm}$ 5.0    &58.6 ${\pm}$ 2.5    &\textbf{55.6 ${\pm}$ 5.4}    &84.4 ${\pm}$ 5.2    &65.6 ${\pm}$ 14.1    & \textless{0.01}  \\
				&EUnet(2023)    &86.4 ${\pm}$ 1.0    &76.9 ${\pm}$ 1.3    &81.7 ${\pm}$ 3.3    &\underline{75.5 ${\pm}$ 1.9}    &\underline{54.1 ${\pm}$ 5.6}    &70.6 ${\pm}$ 7.1    &71.8 ${\pm}$ 13.2    & \textless{0.01}  \\
				&UNesT(2023)    &79.2 ${\pm}$ 1.4    &68.2 ${\pm}$ 1.6    &69.4 ${\pm}$ 4.7    &71.8 ${\pm}$ 2.1    &26.6 ${\pm}$ 4.4    &73.9 ${\pm}$ 5.2    &62.0 ${\pm}$ 24.7    & \textless{0.01}  \\
				&nnUnet(2021)    &78.1 ${\pm}$ 1.0    &\underline{77.2 ${\pm}$ 1.0}    &\textbf{89.3 ${\pm}$ 0.9}    &\textbf{82.0 ${\pm}$ 1.2}    &52.9 ${\pm}$ 5.0    &\textbf{96.7 ${\pm}$ 0.7}    &\textbf{79.6 ${\pm}$ 20.7}    & \textless{0.01}  \\
				\hline
				\multirow{13}{*}{SPE}    &PDPNet    &99.8 ${\pm}$ 0.0    &99.3 ${\pm}$ 0.0    &99.7 ${\pm}$ 0.1    &\textbf{99.1 ${\pm}$ 0.1}    &99.6 ${\pm}$ 0.1    &99.2 ${\pm}$ 0.1    &99.4 ${\pm}$ 0.3    & -  \\
				&UNet3+(2020)    &\textbf{99.9 ${\pm}$ 0.0}    &\underline{99.6 ${\pm}$ 0.0}    &99.6 ${\pm}$ 0.1    &98.9 ${\pm}$ 0.1    &99.4 ${\pm}$ 0.3    &99.4 ${\pm}$ 0.1    &99.4 ${\pm}$ 0.4    & \textless{0.01}  \\
				&UTNet(2021)    &99.8 ${\pm}$ 0.0    &99.5 ${\pm}$ 0.0    &99.7 ${\pm}$ 0.1    &98.9 ${\pm}$ 0.1    &99.5 ${\pm}$ 0.2    &99.6 ${\pm}$ 0.1    &99.4 ${\pm}$ 0.4    & \textless{0.01}  \\
				&DUNet(2020)    &\underline{99.9 ${\pm}$ 0.0}    &\textbf{99.6 ${\pm}$ 0.0}    &99.9 ${\pm}$ 0.0    &98.9 ${\pm}$ 0.1    &99.7 ${\pm}$ 0.1    &99.3 ${\pm}$ 0.1    &99.5 ${\pm}$ 0.5    & \textless{0.01}  \\
				&LEDNet(2019)    &99.9 ${\pm}$ 0.0    &99.5 ${\pm}$ 0.0    &\underline{99.9 ${\pm}$ 0.0}    &99.0 ${\pm}$ 0.1    &\underline{99.7 ${\pm}$ 0.1}    &\underline{99.6 ${\pm}$ 0.1}    &\underline{99.5 ${\pm}$ 0.4}    & \textless{0.01}  \\
				&BiSeNetV2(2021)    &99.8 ${\pm}$ 0.0    &99.4 ${\pm}$ 0.0    &99.9 ${\pm}$ 0.0    &98.3 ${\pm}$ 0.2    &99.2 ${\pm}$ 0.3    &99.3 ${\pm}$ 0.1    &99.2 ${\pm}$ 0.7    & \textless{0.01}  \\
				&AttUnet(2018)    &99.9 ${\pm}$ 0.0    &99.5 ${\pm}$ 0.0    &99.8 ${\pm}$ 0.0    &98.9 ${\pm}$ 0.1    &\textbf{99.8 ${\pm}$ 0.1}    &99.6 ${\pm}$ 0.1    &99.5 ${\pm}$ 0.5    & \textless{0.01}  \\
				&Deeplabv3(2017)    &99.9 ${\pm}$ 0.0    &99.6 ${\pm}$ 0.0    &\textbf{99.9 ${\pm}$ 0.0}    &\underline{99.0 ${\pm}$ 0.1}    &99.5 ${\pm}$ 0.2    &\textbf{99.8 ${\pm}$ 0.1}    &\textbf{99.6 ${\pm}$ 0.4}    & \textless{0.01}  \\
				&AAUnet(2022)    &99.8 ${\pm}$ 0.0    &99.2 ${\pm}$ 0.1    &99.7 ${\pm}$ 0.1    &97.9 ${\pm}$ 0.2    &94.6 ${\pm}$ 0.8    &98.9 ${\pm}$ 0.2    &98.1 ${\pm}$ 2.5    & \textless{0.01}  \\
				&EUnet(2023)    &99.9 ${\pm}$ 0.0    &99.2 ${\pm}$ 0.1    &99.7 ${\pm}$ 0.1    &98.1 ${\pm}$ 0.2    &96.3 ${\pm}$ 0.7    &99.6 ${\pm}$ 0.1    &98.6 ${\pm}$ 1.7    & \textless{0.01}  \\
				&UNesT(2023)    &99.8 ${\pm}$ 0.0    &99.4 ${\pm}$ 0.1    &99.7 ${\pm}$ 0.1    &98.1 ${\pm}$ 0.2    &98.8 ${\pm}$ 0.3    &99.5 ${\pm}$ 0.1    &99.1 ${\pm}$ 0.8    & \textless{0.01}  \\
				&nnUnet(2021)    &99.9 ${\pm}$ 0.0    &99.2 ${\pm}$ 0.1    &99.2 ${\pm}$ 0.2    &98.2 ${\pm}$ 0.2    &97.8 ${\pm}$ 0.4    &97.7 ${\pm}$ 0.2    &98.4 ${\pm}$ 0.9    & \textless{0.01}  \\
				\hline
				\multirow{13}{*}{HD95}    &PDPNet    &\textbf{2.40 ${\pm}$ 0.27}    &\textbf{9.16 ${\pm}$ 0.62}    &10.56 ${\pm}$ 2.44    &\textbf{12.11 ${\pm}$ 0.80}    &22.74 ${\pm}$ 2.87    &\textbf{6.37 ${\pm}$ 0.85}    &\textbf{12.18 ${\pm}$ 7.78}    & -  \\
				&UNet3+(2020)    &3.16 ${\pm}$ 0.38    &\underline{9.69 ${\pm}$ 0.63}    &17.34 ${\pm}$ 3.21    &14.25 ${\pm}$ 0.85    &22.74 ${\pm}$ 2.13    &10.31 ${\pm}$ 1.45    &14.87 ${\pm}$ 6.69    & \textless{0.01}  \\
				&UTNet(2021)    &4.97 ${\pm}$ 0.49    &10.97 ${\pm}$ 0.72    &8.45 ${\pm}$ 1.80    &14.68 ${\pm}$ 0.82    &23.18 ${\pm}$ 2.05    &8.69 ${\pm}$ 1.12    &13.19 ${\pm}$ 7.59    & \textless{0.01}  \\
				&DUNet(2020)    &\underline{2.96 ${\pm}$ 0.35}    &11.15 ${\pm}$ 0.74    &\textbf{6.95 ${\pm}$ 1.41}    &13.58 ${\pm}$ 0.84    &26.22 ${\pm}$ 2.09    &9.59 ${\pm}$ 1.33    &13.50 ${\pm}$ 9.32    & \textless{0.01}  \\
				&LEDNet(2019)    &3.69 ${\pm}$ 0.34    &9.94 ${\pm}$ 0.65    &9.55 ${\pm}$ 1.82    &16.79 ${\pm}$ 0.86    &27.44 ${\pm}$ 1.83    &10.83 ${\pm}$ 1.42    &14.91 ${\pm}$ 9.43    & \textless{0.01}  \\
				&BiSeNetV2(2021)    &3.78 ${\pm}$ 0.35    &10.29 ${\pm}$ 0.65    &11.19 ${\pm}$ 2.02    &13.51 ${\pm}$ 0.79    &\underline{22.31 ${\pm}$ 2.23}    &9.52 ${\pm}$ 1.42    &13.36 ${\pm}$ 6.48    & \textless{0.01}  \\
				&AttUnet(2018)    &4.77 ${\pm}$ 0.46    &10.52 ${\pm}$ 0.67    &11.88 ${\pm}$ 2.09    &\underline{13.19 ${\pm}$ 0.77}    &\textbf{22.14 ${\pm}$ 2.04}    &8.67 ${\pm}$ 1.23    &13.28 ${\pm}$ 6.50    & \textless{0.01}  \\
				&Deeplabv3(2017)    &4.21 ${\pm}$ 0.37    &9.84 ${\pm}$ 0.61    &\underline{7.55 ${\pm}$ 1.49}    &15.00 ${\pm}$ 0.86    &25.51 ${\pm}$ 2.39    &\underline{8.05 ${\pm}$ 1.15}    &\underline{13.19 ${\pm}$ 9.30}    & \textless{0.01}  \\
				&AAUnet(2022)    &6.29 ${\pm}$ 0.54    &12.86 ${\pm}$ 0.69    &10.47 ${\pm}$ 1.82    &18.77 ${\pm}$ 0.92    &29.62 ${\pm}$ 1.89    &12.98 ${\pm}$ 1.85    &16.94 ${\pm}$ 9.58    & \textless{0.01}  \\
				&EUnet(2023)    &3.83 ${\pm}$ 0.40    &10.64 ${\pm}$ 0.66    &12.25 ${\pm}$ 2.99    &14.85 ${\pm}$ 0.90    &29.07 ${\pm}$ 2.51    &8.80 ${\pm}$ 1.17    &15.12 ${\pm}$ 10.06    & \textless{0.01}  \\
				&UNesT(2023)    &5.77 ${\pm}$ 0.49    &10.26 ${\pm}$ 0.55    &10.57 ${\pm}$ 2.01    &15.53 ${\pm}$ 0.87    &25.92 ${\pm}$ 2.42    &12.83 ${\pm}$ 1.77    &15.02 ${\pm}$ 8.01    & \textless{0.01}  \\
				&nnUnet(2021)    &5.00 ${\pm}$ 0.44    &13.80 ${\pm}$ 0.77    &25.24 ${\pm}$ 4.04    &16.37 ${\pm}$ 0.92    &27.73 ${\pm}$ 2.68    &24.06 ${\pm}$ 2.20    &21.44 ${\pm}$ 7.48    & \textless{0.01}  \\
				\hline
			\end{tabular}
		}
		\begin{tablenotes}
			\footnotesize
			\item [*] The optimal metrics are highlighted in bold, and suboptimal metrics are underlined.
		\end{tablenotes}
	\end{threeparttable}
\end{table*}

Table II also compares the quantitative metrics of each method on both validation and multi-center test sets. It can be seen that our method achieves the best performance in terms of DSC, SEN and HD95 on validation set. Compared against the corresponding suboptimal methods, the DSC, SEN and HD95 are improved by 0.44\%, 4.86\% and 20.0\%, respectively. On the multi-center test sets, the proposed method outperforms the others in terms of average DSC and HD95 (the last column), they are improved at least by 5.13\% and 7.58\%, respectively. Regarding the SEN, nnUnet obtains the highest average SEN of 79.6\% but sacrifices the average HD95 which is almost up to 21.4. Even though the SEN of our PDPNet is a little (about 7.29\%) inferior to that of nnUnet, its HD95 is about 43.0\% better than that of nnUnet. As for the SPE, no matter on validation or multi-center test sets, all the models can get a SPE higher than 97\%, and there is no significant difference in SPE among different methods. From the prospective of the performance of different models on each test cohort, we notice that the samples in GPH and HUM-S cohorts are easy to segment, while the samples in two publicly available cohorts (ISPY-G and ISPY-S) and TPH cohort are difficult to segment. On GPH dataset, our method obtains the best performance in terms of all the metrics (except for SPE which is 99.8\%, 0.1\% lower than that of UNet3+). On HUM-S dataset, although our method did not achieve the best performance, it can balance different evaluation metrics well, unlike other methods that are only good on one metric and poor on others, such as EUnet which is good on DSC but poor on SEN, and nnUNet is good on SEN while poor on HD95. On ISPY-G, ISPY-S and TPH datasets, the performance of the most models is not satisfactory, the DSC of several models, such as DUnet and  LEDNet, is even lower than 20\%. Generally, our proposed method and nnUNet obtain the comparable results. That means, on  ISPY-G and ISPY-S cohorts, our method achieves a little lower but comparable DSC with nnUnet, while on THP cohort, our DSC is much higher (about 32.5\%) than that of nnUnet. Even though the SEN of nnUnet is better than that of our method on these three cohorts, its HD95 values are much higher than ours.
\begin{figure}[htbp]
	\centering
	\vspace{0cm}
	\setlength{\abovecaptionskip}{0cm}
	\setlength{\belowcaptionskip}{0cm}
	\subfigure[Validation set]{
		\setlength{\abovecaptionskip}{0cm}
		\includegraphics[width=0.4\textwidth]{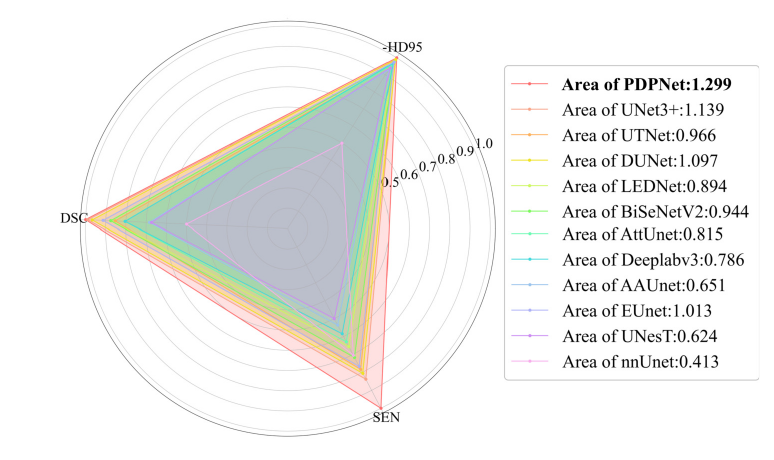}
		\label{fig:4a}
	}
	\subfigure[Multi-center test sets]{
		\setlength{\abovecaptionskip}{0cm}
		\includegraphics[width=0.4\textwidth]{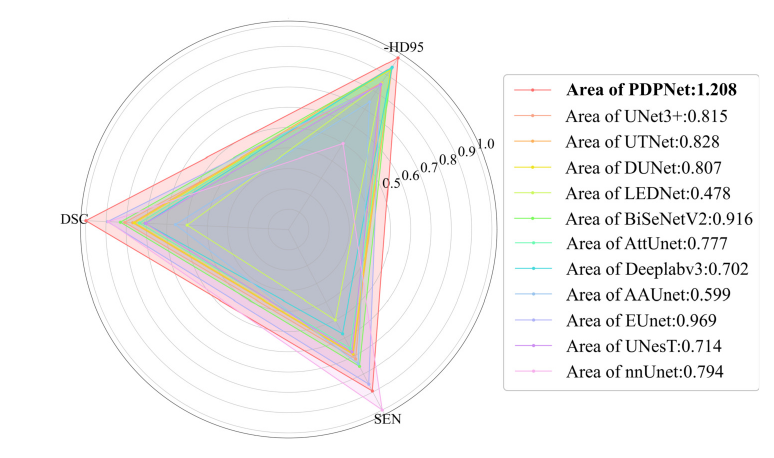}
		\label{fig:4b}
	}
	\captionsetup{justification=justified}
	\caption{Radar plots of average evaluation metrics obtained with different methods on both validation (a) and multi-center test sets (b). The number indicated in each legend represents the surface of area enclosed by three metrics. The bigger area indicates the better performance.}
	\label{fig:4}
\end{figure}

To comprehensively compare different methods, Fig.\ref{fig:4} shows the radar plots of all the metrics (except SPE since it does not almost change) obtained by each method on both validation (a) and multi-center test sets (b). Since smaller HD95 is better, negative values of HD95 are used in the radar plots for ease of comparison. Subsequently, all metrics are normalized between 0.5 and 1 to emphasize the differences between the methods. Moreover, the number indicated in the legends represents the area surrounded by three metrics. From these radar plots, it can be clearly observed that no matter on which sets, the radar plots associated with our method (PDPNet, red color) covers the largest area, indicating the best overall performance when considering DSC, HD95, and SEN simultaneously.

\begin{figure}[htb]
	\centering
	\vspace{0cm}
	\setlength{\abovecaptionskip}{0cm}
	\setlength{\belowcaptionskip}{0cm}
	\subfigure[Curves of DSC]{
		\setlength{\abovecaptionskip}{0cm}
		\includegraphics[width=0.4\textwidth]{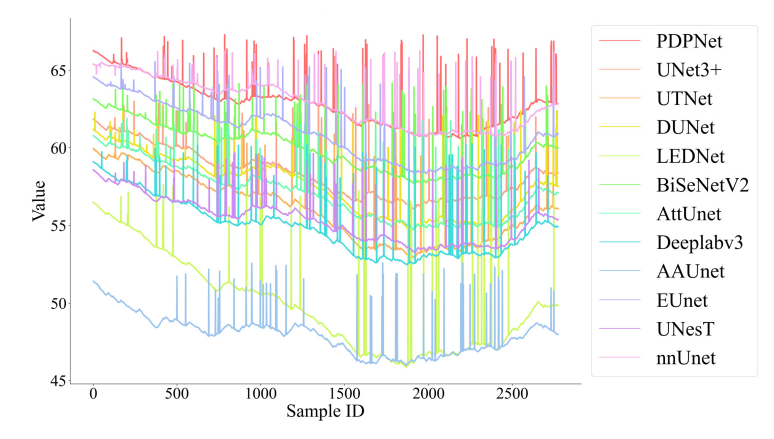}
		\label{fig:5a}
	}
	\subfigure[Curves of HD95]{
		\setlength{\abovecaptionskip}{0cm}
		\includegraphics[width=0.4\textwidth]{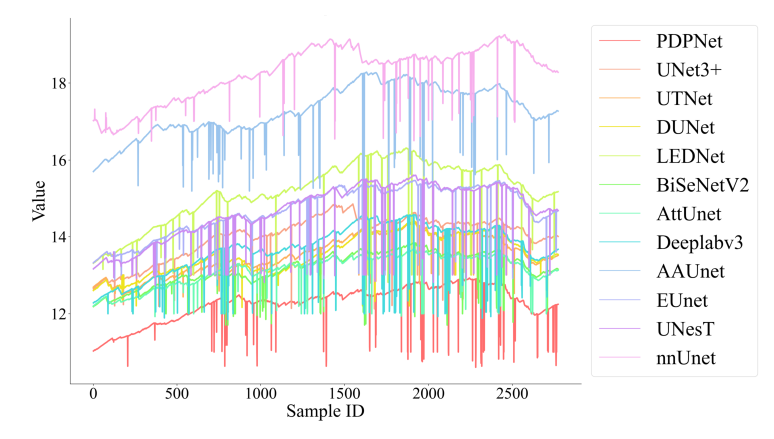}
		\label{fig:5b}
	}
	\caption{The curves of average DSC for all the samples in five test cohorts obtained by different methods.}
	\label{fig:5}
\end{figure}

To further more intuitively evaluate the superiority of the proposed method on the multi-center test sets, the curves of average DSC and HD95 (within 95\% confidence interval) for all the samples in five cohorts are plotted in Fig. \ref{fig:5}. It can be seen that, in terms of DSC, our method and nnUnet are much better than the others, for most samples, the DSC of our method is a little higher than that of nnUnet. As for HD95, our PDPNet obtains the lowest HD95 on all the samples while nnUnet yields the highest ones.

\subsection{Ablation results}
\textit{ 1) Ablation in network modules}

\begin{figure}[H]
	\centering
	\vspace{0cm}
	\setlength{\abovecaptionskip}{0cm}
	\setlength{\belowcaptionskip}{0cm}
	\includegraphics[width=0.4\textwidth]{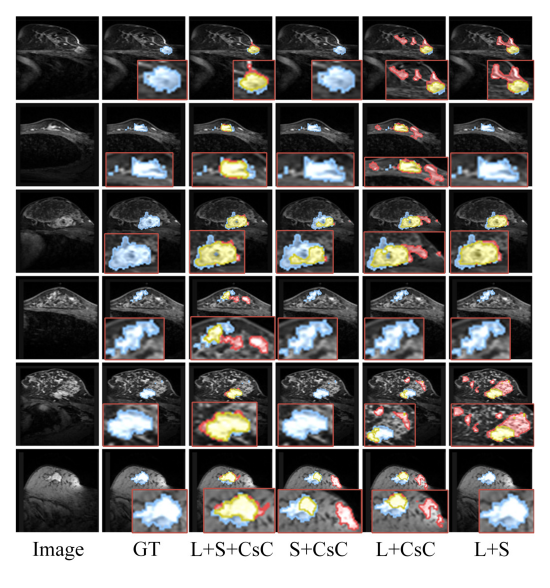}
	\captionsetup{justification=justified}
	\caption{Segmentation results of different ablation experiments on one test set (GPH cohort). GT signifies the ground truth, L+S+CsC means the proposed model with localization module (L), weak semantic prior (S) and cross scale correlation prior (CsC); S+CsC represents the model without using localization module; L+CsC is the model without using weak semantic prior; L+S indicates the model without using cross scale correlation prior.}
	\label{fig:6}
\end{figure}

In order to verify the effectiveness of the proposed localization module (shorten as L), weak semantic prior (shorten as S), as well as the cross-scale correlation prior (shorten as CsC), we remove them respectively from the network. The segmentation results for different ablation experiments are shown in Fig \ref{fig:6}. Note that, without the localization module (S+CsC), the model cannot distinguish the small tumors (the first two rows), and is easy to generate false background for tumors with irregular shapes (the 3$^{rd}$ and 4$^{th}$ rows) and similar intensity to the normal tissues (the last two rows). Without using the semantic prior (L+CsC), the model predicts more false backgrounds (blue regions), while without using cross-scale correlation prior (L+S), the model cannot distinguish the tumors with similar intensity to the background.

\begin{figure}[htbp]
	\centering
	\vspace{0cm}
	\setlength{\abovecaptionskip}{0cm}
	\setlength{\belowcaptionskip}{0cm}
	\includegraphics[width=0.8\linewidth]{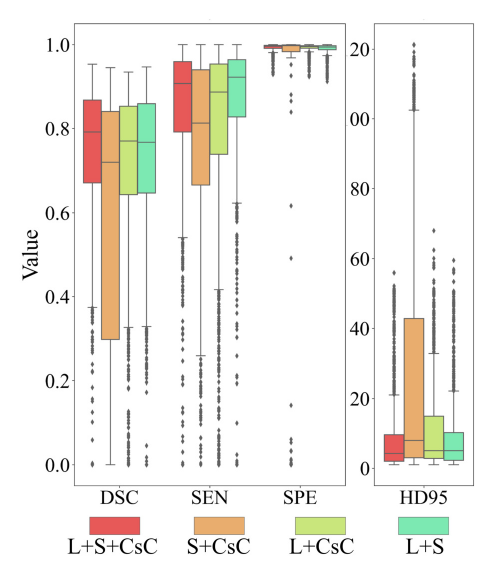}
	\captionsetup{justification=justified}
	\caption{Box plots of different evaluations metrics for ablation results on one test set (GPH cohort). L+S+CsC is the proposed model, S+CsC is the model without using localization module, L+CsC is the model without using weak semantic prior and L+S is the model without using cross scale correlation prior.}
	\label{fig:7}
\end{figure}

To quantitatively show the effects of the localization module and dual priors, the box plots of evaluation metrics obtained with different ablation experiments are drawn in Fig. \ref{fig:7}. We find that removing the localization module from the PDPNet (S+CsC), the median DSC and SEN are decreased by 9.10\% and 10.36\% respectively, and HD95 is increased by 87.4\%. When removing the semantic prior module (L+CsC), the median DSC, SEN and SPE decrease slightly (2.65\%,  2.21\% and 0.1\% respectively), and HD95 is increased by 17.8\%. Without the cross-scale correlation prior (L+S), the median DSC and SPE are decreased by 3.03 \% and 0.20\%, the median HD95 is increased by 17.80\%, and the SEN is increased by 1.76\%. In addition, with the help of localization module and dual priors (L+S+CsC), the interquartile ranges of all the evaluation metrics are the smallest, and the second and third quartiles are also the highest, which indicates that the localization module and dual priors can assist the model to achieve the best performance for the most of samples.

\textit{ 3) Ablation in dual prior settings}

\begin{figure}[htb]
	\centering
	\vspace{0cm}
	\setlength{\abovecaptionskip}{0cm}
	\setlength{\belowcaptionskip}{0cm}
	\includegraphics[scale=0.6]{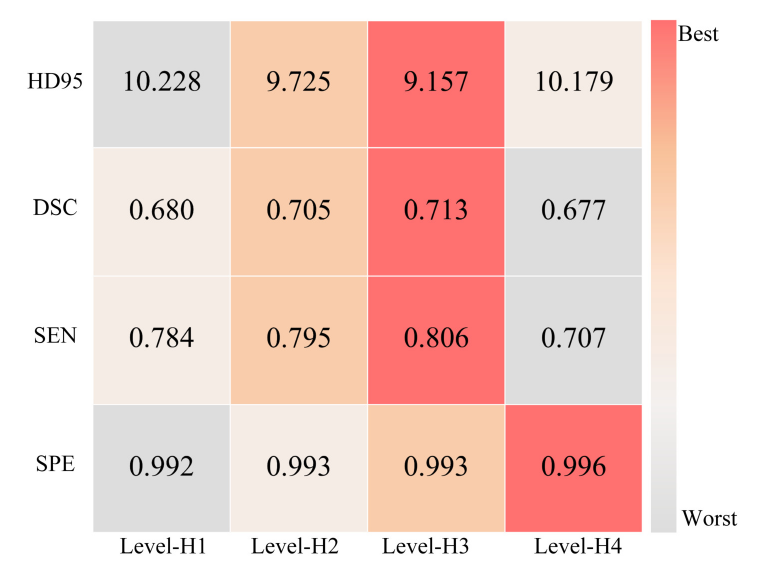}
	\captionsetup{justification=justified}
	\caption{Heatmap of evaluation metrics obtained with different dual prior settings on one test set (GPH cohort). ''Level-H1'' means using dual prior module (DPM) only between level 5 and level 4; ''Level-H2'' indicates using the DPM between level 4 and level 3 in addition to ''Level-H1''; ''Level-H3'' means using DPM between level 3 and level 2 besides ''Level-H1'' and ''Level-H2''; ''Level-H4'' indicates using DPM between all the levels, including between level 2 and level 1.}
	\label{fig:8}
\end{figure}

Through the module ablation experimental results, we have verified that the DPM is beneficial for improving the segmentation performance of the model. To further demonstrate how to use the DPM is the most effective, we test four different DPM settings: Level-H1, introducing the DPM only between level 5 and level 4 (the levels are indicated in the Fig. \ref {fig:2}); Level-H2, introducing the DPM between level 5 and level 4, as well as between level 4 and level 3; Level-H3, introducing DPM between level 4 and level 5, level 4 and level 3, as well as level 3 and level 2 (Level-H3); Level-H4, based on Level-H3, introducing also the DPM between level 2 and level 1, which amounts to inferring $F_1$ with $P_2$, $M_1$ and $F_2$ (not illustrated in Fig. \ref {fig:2}). The evaluation metrics obtained with different settings are illustrated in the heatmap of Fig. \ref{fig:8}. It can be observed that, with the DPM setting of Level-H3, the model obtains the best DSC, SEN, and HD95, while with the setting of Level-H4, the model performance decreases.

\section{Discussion}
\label{sec:Discussion}
To improve the generalization ability of breast tumor segmentation models and to increase the segmentation accuracy for tumors with irregular shapes and similar intensity to normal tissues, we proposed a PDPNet to segment the breast tumors acquired from different centers. It used a progressive strategy and dual priors to promote the segmentation performance. More precisely, we first localized the tumors with the corse segmentation strategy and then cropped the images based on the localization to allow the model focus only on the tumor regions. Subsequently, we introduced the cross-scale correlation prior and weak semantic prior into the progressive inference of segmentation maps. With the help of these two priors, feature representation ability about the tumors can be further enhanced, thus improving the generalization performance of the model. Through the comparisons against the SOTA methods on six cohorts from different centers and the ablation experiments, we validated that the proposed progressive strategy and dual priori knowledge based segmentation network are useful for delineating multi-center breast tumors. 

\begin{figure}[htbp]
	\centering
	\setlength{\abovecaptionskip}{0cm}
	\includegraphics[width=0.9\linewidth]{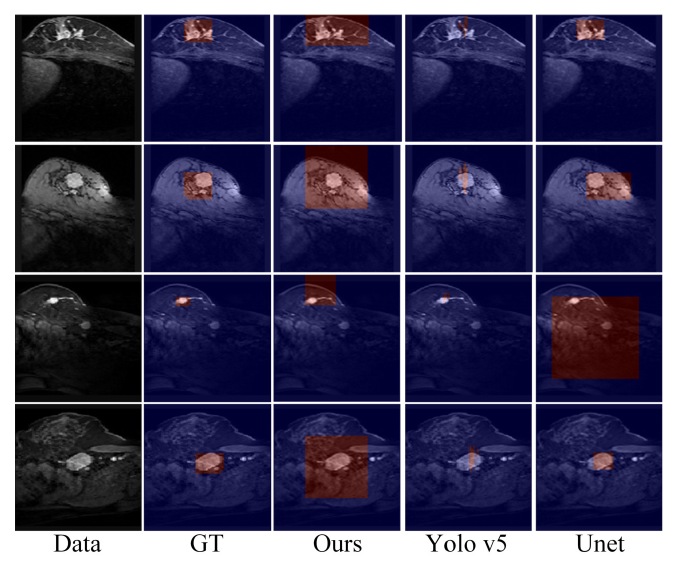}
	\label{fig:30}
	\captionsetup{justification=justified}
	\caption{Comparing our localization module against the commonly used localization methods.}
	\label{fig9}
\end{figure}

From the comparison results (Fig. \ref{fig:3}), we found that even for the validation set (coming from the same cohort with the train set), the existing methods cannot well segment the tumors with smaller size, irregular shapes and lower intensity contrast (similar intensity to normal tissues), which indicates that the deep learning based segmentation models are sensitive to the low-level image features, such as image intensity distribution and shapes of the tumors. However, with the help of the proposed localization strategy and dual priors, segmentation performance is significantly improved on several test sets. The reason is that the proposed localization module based on coarse segmentation (weak semantic prior) can roughly provide the location information of the tumors, using which to crop the tumor regions can depress the influence of some normal tissues on the subsequent segmentation, thus allowing the model to better segment tumors with smaller size and lower contrast. Currently, employing the localization model to first crop the ROI is a common strategy to refine the segmentation result, mainly including the object-detection based localization and coarse segmentation-based localization methods. The former takes the original image as input, and the bounding box of tumor region as the learning target, the tumor localization can be realized by regressing the height, width and center of bounding box. The latter uses a network to implement the coarse segmentation, and then uses the bounding box of segmentation results to localize the tumor region. Both of them are easily influenced by the image contrast or image textures, therefore, it is difficult to achieve the satisfied localization performance, especially for multi-center datasets. This can be reflected by Fig. \ref{fig9}, where the localization results derived from our localization module and those obtained with object detection method YOLOV5 and coarse segmentation method UNet are compared.

It can be seen that YOLOv5 can find the location of the tumor center, but it cannot regress accurately the bounding box of the tumor. Regarding the localization results derived from UNet, due to the influence of false tumors (the third row) or false backgrounds (the last row), the localization region becomes too large or too small. In contrast, in this work, we used the high-level semantic features to predict whether the corresponding image patch contains the tumor, rather than using the original image to predict more complex tasks, such as regression (width and height) or dense prediction (segmentation). Such simplified patch-classification task is not sensitive to the image details and therefore is able to achieve a better localization performance (Fig. \ref{fig9}). In addition, different from the existing localization-assisted segmentation models, our localization module and the refine segmentation network are trained simultaneously rather than in a two-stage learning manner, this cooperative training is beneficial for mutual promotion.

\begin{figure}[htbp]
	\centering
	\vspace{0cm}
	\setlength{\abovecaptionskip}{0cm}
	\setlength{\belowcaptionskip}{0cm}
	\subfigure[Difference in intensity distributions between low-contrast tumors before (Raw) and after (Cropped) localization ]{
		\setlength{\abovecaptionskip}{0cm}
		\includegraphics[width=0.45\textwidth]{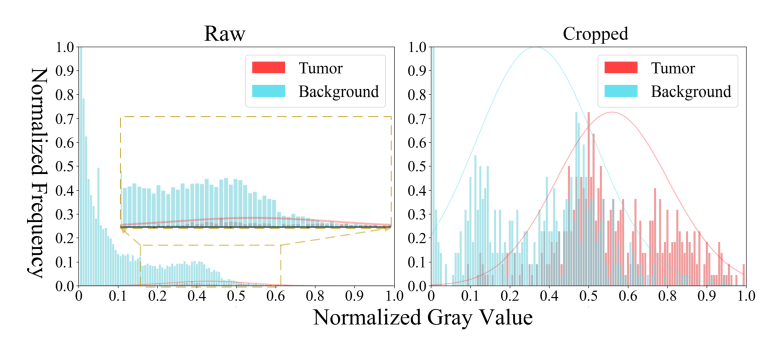}
		\label{fig:9a}
	}
	\subfigure[Difference in intensity distributions between small tumors and background before (Raw) and after (Cropped) localization.]{
		\setlength{\abovecaptionskip}{0cm}
		\includegraphics[width=0.45\textwidth]{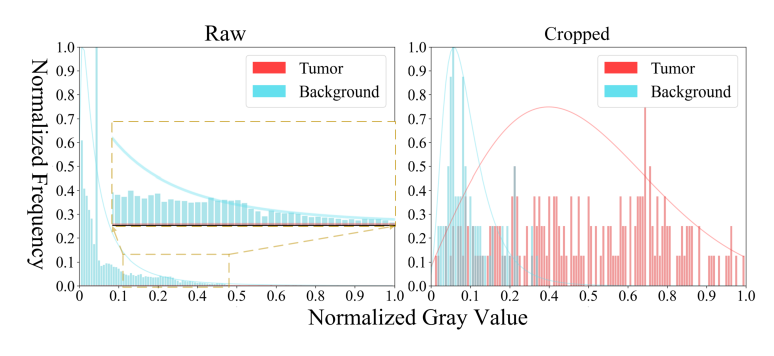}
		\label{fig:9b}
	}
	\captionsetup{justification=justified}
	\caption{Differences in intensity distributions before and after localization. Red and cyan represent different individual distributions, and gray indicates the overlapped distribution between them.}
	\label{fig:9}
\end{figure}

The effectiveness of our localization module can also been validated by the ablation results. As illustrated in ``S+CsC'' of Fig. \ref{fig:6}, without localization module,  when the tumor region is too small or the intensity difference between the tumors and normal tissues is not obvious, the model cannot discriminate the semantic features of background from those of tumors. However, if cropping the images using localization module, the background information is limited and its influence can be overlooked. As illustrated in Fig. \ref{fig:9a} and Fig. \ref{fig:9b}, before the localization (Raw), the intensity distributions of tumors with low-contrast (Fig. \ref{fig:9a}) or smaller size (Fig. \ref{fig:9b}) are overlapped or submerged by those of background, while after the localization (Cropped), the intensity distribution of tumors and the remaining background can be distinguished.

\begin{figure*}[htb]
	\centering
	\vspace{0cm}
	\setlength{\abovecaptionskip}{0cm}
	\setlength{\belowcaptionskip}{0cm}
	\subfigure[Cumulative probability distribution of intensity difference between tumor regions and normal tissues in different cohorts]{
		\setlength{\abovecaptionskip}{0cm}
		\includegraphics[width=0.45\textwidth]{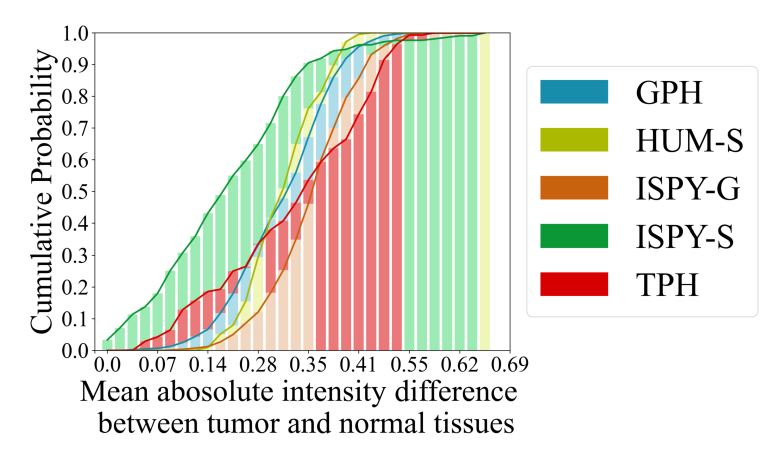}
		\label{fig:10a}
	}
	\subfigure[Cumulative probability distribution of the ratio of tumor size to image size in different cohorts]{
		\setlength{\abovecaptionskip}{0cm}
		\includegraphics[width=0.45\textwidth]{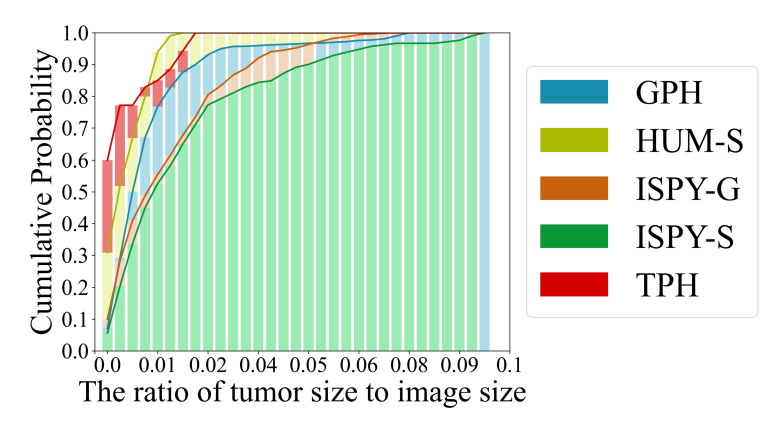}
		\label{fig:10b}
	}
	\captionsetup{justification=justified}
	\caption{Distributions of intensity difference between tumor regions and normal tissues (a), as well as the ratio of tumor size ratio image size in different test cohorts (b).}
	\label{fig:10}
\end{figure*}

In addition, from Table II we notice that almost all the models perform the best on HUM-S cohort, following by GPH cohort and then TPH cohort. But on two publicly available cohorts ISPY-G and ISPY-S, their performance is not satisfactory. To explore the reasons, in Fig.\ref{fig:10} shows the distribution of intensity difference between tumor regions and normal tissues in different cohorts (Fig.\ref{fig:10a}),  as well as the distribution of the ratio of tumor size to the image size (Fig.\ref{fig:10b}). We observe that, the tumors in HUM-S, GPH and TPH cohorts have the smaller size (the cumulative probability of tumor size ratio approaches to 1 quickly in Fig.\ref{fig:10b}), while the tumor size in ISPY-G and ISPY-S cohorts varies a lot (the cumulative probability increases slowly to 1). From the visualization results of Fig.\ref{fig:3}, we know that the wrong segmentations usually happen at the regions with extremely high or low intensity. Due to the application of the proposed localization module, the interference of the extremely low or high intensity outside the tumor can be mostly eliminated in our method. While inside the large tumors, it has more chance to have heterogeneous intensity distribution (including extremely high or low intensity) than inside small tumors, which results in some regions inside large tumors being wrongly segmented. Accordingly, our method performs better on cohorts with smaller tumor size (GPH, HUM-S and TPH) than those with larger tumor size (ISPY-G and ISPY-S). Moreover, from Fig.\ref{fig:10a}, we also notice that intensity difference between tumor region and background in ISPY-S cohort is so small, making it difficult to distinguish tumors from background, which therefore degrades further the segmentation performance of all models on this cohort. This is why most of the model achieve the worst performance on ISPY-S cohort. Only our PDPNet and nnUnet can obtain good comparable performance. As we mentioned above, since our model cannot address the influence of heterogenous intensity inside the tumor on the segmentation results, it may mis-segment some regions in the tumor and therefore generate low SEN. In contrast, nnUnet uses the overlapped sliding patches and test time augmentation (TTA) strategy for inference, which can deal with the mis-segmentation issue to some extent, therefore yielding a higher SEN than our method. However, since nnUnet was trained with image patches, some important context information may be lost, which results in therefore some false tumors (the regions with extremely high intensity are taken as tumors, red regions in Fig.3), leading to the increase of HD95. While in the proposed PDPNet, the weak semantic prior and cross-scale correlation prior are beneficial for retaining the shape and boundary information of the tumors, decreasing therefore HD95. This is why the HD95 of our model is much lower than that of nnUnet.

\begin{figure*}[htb]
	\centering
	\vspace{0cm}
	\setlength{\abovecaptionskip}{0cm}
	\setlength{\belowcaptionskip}{0cm}
	\includegraphics[width=0.9\textwidth]{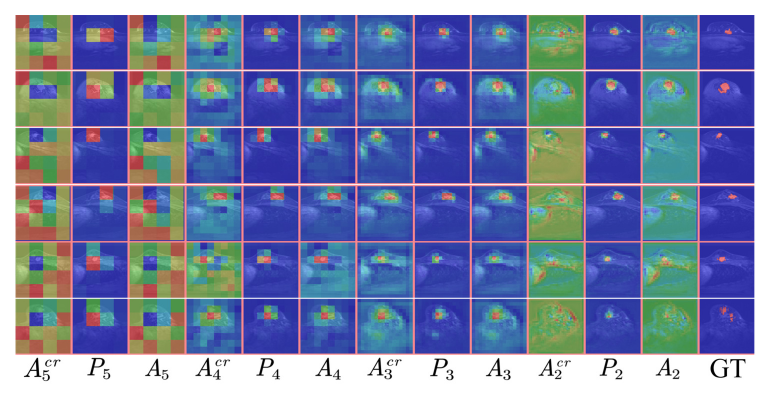}
	\captionsetup{justification=justified}
	\caption{Visualization of prior maps for several samples in multi-center test sets. $P_s$, $A_s^{cr}$ and $A_s=P_s+A_s^{cr}$ represent weak semantic prior maps, cross scale correlation prior maps and dual prior maps respectively at different levels, with $s=2,3,4,5$. Each row shows the prior maps of one sample.}
	\label{fig:11}                 
\end{figure*}

Even though the dual prior modules can promote the performance, it does not mean that introducing the dual priors at all the scales can get the best results. As illustrated in Fig. \ref{fig:8}, at the high-level semantic scales (from level 5 to level 2 in Fig. \ref{fig:2}), introducing the dual priors was useful for improving the segmentation performance, with the DSC, SEN, and HD95 being improved progressively from level-H1 to level-H3. However, adding the dual prior modules in the low-level semantic scale (between level 2 and level 1 in Fig. \ref{fig:2}), led to a dramatic decrease in almost all the evaluation metrics (as illustrated in the Level-H4 in Fig.\ref{fig:8}). The reason may be that at the low-level semantic scale, the cross-scale correlation prior rather than weak semantic prior plays the dominant role, resulting in that the model is not aware of where should the attention to pay. To provide further evidence, we interpolated the weak semantic priors $P_s$ ($s=2,3,4,5$), the cross-scale correlation prior maps ($A_s^{cr}$, calculated with Eq. \ref {eq:8}) and the dual prior maps ($A_s$, adding $P_s$ and $A_s^{cr}$) to the same size as the input image, and normalized their values to the range of [0,1], as shown in Fig. \ref{fig:11}. Comparing with the ground truth of the tumor regions (last column), we found that the large values at the low-resolution weak semantic prior maps ($P_5$ and $P4$) were mainly concentrated on the tumor region, which indicates that the weak semantic priors can indeed assist the model in tumor localization and thus improve the performance of the model. Cross-scale correlation prior maps $A_s^{cr}$ had high values in the tumors and surround regions, making the model consider not only the tumor but also the surrounding regions, accordingly promoting the shape-aware ability of the model. Combing the semantic prior and cross-scale correlation prior can provide complementary information to further improve the segmentation performance, as illustrated in the maps $A_4$ and $A_3$, where the tumor location and the tumor details can be clearly observed. However, once the DPM were implemented on low levels, that means calculating the cross-scale correlation prior $A_2^{cr}$ with low-level semantic features $M_1$ and high level semantic features $F_2$, we notice that $A_2^{cr}$ moves away from the ground truth (comparing $A_2^{cr}$ and GT), introducing a lot of wrongly highlighted regions (in red). Even though the weak semantic prior at this level ($P_2$) is close to GT, due to the influence of cross scale correlation prior $A_2^{cr}$, adding correlation prior and weak semantic prior together still generates biased dual prior map ($A_2$). Accordingly, using such biased prior maps to guide the segmentation will result in worse performance, this is why using the DPM with the setting of ``Level-H4'' generates worse performance than ``Level-H3'', and also the reason why we did not involve $M_1$ in our segmentation network (as illustrated in Fig.\ref{fig:2}).

Although all the experiments have shown the superiority of the proposed PDPNet, it still has certain limitations. First, due to the limited samples in the training set, the feature representations of the breast tumors may be biased, which prohibits the model from accurately recognizing the tumor regions with heterogeneous intensity, such as on ISPY-S set, this needs to be addressed in the future. Moreover, in the localization module, we used a handcrafted threshold to annotate the patch label, which may miss some tumor information, even though using the data augmentation strategies can deal with such issue to some extent, how to effectively and accurately annotate the patch label is also of interest. Furthermore, as we demonstrated in the whole manuscript, the cross correlation prior proposed in this work has the benefit of sensing the tumor shapes, but it presents bias at low semantic level, how to design another shape prior that allows to use the features at all levels may further improve the segmentation performance. Finally, although the proposed method achieves a better generalization performance on 2D segmentation, how to fully explore the information of DCE-MRI and extend the model to 3D or even 4D to further improve the generalization ability still requires much efforts.

\section{Conclusion}
\label{sec:Conclusion}
A progressive dual priori network for multi-center breast tumor segmentation was proposed. It consisted of a localization modules and a dual prior knowledge-based segmentation module. The localization module was realized using patch-level segmentation, and the segmentation module refined progressively the tumor masks using multi-scale weak semantic priors and cross-scale correlation priors. The comparison and ablation results showed that the proposed method outperformed the SOTA methods on both validation and multi-center test sets, and that the proposed progressive strategy and the dual priors are useful for promoting the generalization ability of the segmentation models. 

\bibliographystyle{IEEEtran}
\bibliography{Reference}

\end{document}